\documentclass{article}

\usepackage{spconf,amsmath,graphicx,amsfonts}
\usepackage{enumitem}
\usepackage{hyperref}
\usepackage{etoolbox}
\usepackage{tabularx}
\patchcmd{\thebibliography}
  {\settowidth}
  {\setlength{\itemsep}{0pt plus 0.1pt}\settowidth}
  {}{}
\apptocmd{\thebibliography}
  {\small}
  {}{}
  
\DeclareMathOperator{\PEG}    {PE_G}
\DeclareMathOperator{\PE}    {PE}
\newcommand{\G}{{G}}
\newcommand{\X}{\mathbf{X}} 
\newcommand{\Na}{\mathbb{N}}
\newcommand{\Nm}{\mathbb{N}_m}
\newcommand{\R}{\mathbb{R}} 
\newcommand{\A}{\mathbf{A}}
\newcommand{\Y}{\mathbf{Y}} 

\newcommand{\V}{\mathcal{V}}
\newcommand{\E}{\mathcal{E}}  
\newcommand{\card}[1]{\lvert#1\rvert}
\DeclareMathOperator{\MIX}    {MIX}
\DeclareMathOperator{\row}    {row}

\DeclareMathOperator{\RGG}    {RGG}

\usepackage{tikz}
\usepackage{subcaption} 
\usepackage[font=small,labelfont=bf]{caption}
\title{Graph-based permutation patterns for the analysis of task-related fMRI signals on DTI networks in mild cognitive impairment}
\name{John S.~Fabila-Carrasco\sthanks{These authors share first authorship. This work was supported by the Leverhulme Trust via a Research Project Grant (RPG-2020-158) to JER and by Alzheimer’s Society Grants AS-R42303 and AS-SF-14-008 awarded to MAP in collaboration with JER. ACC acknowledges Edinburgh University's Principle's Career Development PhD Scholarship and Federica Guazzo for pre-processing the fMRI data. For the purpose of open access, the author has applied a Creative Commons Attribution (CC BY) licence to any author accepted manuscript version arising from this submission.}$^\dagger$, Avalon Campbell-Cousins$^{*\dagger}$, Mario A.~Parra-Rodriguez$^{+}$, Javier Escudero$^{\dagger}$ }
\address{$^\dagger$ School of Engineering, IDCOM, University of Edinburgh, UK \\
$^+$ Department of Psychological Sciences and Health, University of Strathclyde, UK}

\begin{document}

\maketitle

\begin{abstract}
Permutation Entropy ($\PE$) is a powerful nonlinear analysis technique for univariate time series. Recently, Permutation Entropy for Graph signals ($\PEG$) has been proposed to extend PE to data residing on irregular domains. However, $\PEG$ is limited as it provides a single value to characterise a whole graph signal. Here, we introduce a novel approach to evaluate graph signals \emph{at the vertex level}: graph-based permutation patterns. Synthetic datasets show the efficacy of our method. We reveal that dynamics in graph signals, undetectable with $\PEG$, can be discerned using our graph-based patterns. These are then validated in DTI and fMRI data acquired during a working memory task in mild cognitive impairment, where we explore functional brain signals on structural white matter networks. Our findings suggest that graph-based permutation patterns in individual brain regions change as the disease progresses, demonstrating potential as a method of analyzing graph-signals at a granular scale. 
\end{abstract}


\section{Introduction}
\label{sec:intro}

Entropy-based nonlinear analysis techniques have become particularly valuable for analysing noisy or short time series related to complex systems~\cite{azami2023entropy,ribeiro2021entropy}. These methods offer insights into signal irregularity, revealing effects such as financial crisis in time series~\cite{yin2014weighted} and anomalies in mechanical and physiological systems~\cite{azami2016improved}. Among them, permutation entropy ($\PE$) is noted for its robustness to noise, fast calculation, and sound statistical properties~\cite{cao2004detecting}.

Building on Shannon's entropy, $\PE$ quantifies the distribution of `permutation patterns' in time series~\cite{Bandt2002}. Such patterns have broad applications, from biomedical to finance data~\cite{riedl2013practical}. Advanced quantifiers further refine time series analysis, like `forbidden patterns' in finance~\cite{Zunino2009} and weighted differences of pattern likelihoods for univariate brain signals~\cite{Bandt2023}. Additionally, the contrasts $\alpha$ (turning rate) and $\beta$ (up-down balance) further refine permutation pattern analysis~\cite{Bandt2023}. While $\PE$ is powerful, its univariate focus is a limitation. A multivariate $\PE$ version exists but it dilutes individual channel characteristics~\cite{morabito2012multivariate}.  More recently, a 2D version of $\PE$ has been proposed for images~\cite{Morel2021}.

Graph signals offer a novel avenue for data analysis on irregular domains~\cite{dong2020graph, ortega2018graph}. The framework of graph signals is highly relevant for a wide variety of settings, such as weather patterns or vehicular traffic~\cite{ortega2022introduction}. One particularly relevant example is neuroimaging, where brain activity can naturally be seen as a graph signal measured over a brain network~\cite{bessadok2022graph, huang2018graph, li2021graph}. In this context, we have recently introduced $\PEG$ for graph signals~\cite{fabila2022permutation}, extending $\PE$ to irregularly sampled data. 

Permutation patterns have received considerable attention recently due to their useful properties in univariate time series, and their study has very recently been extended to 2D formulations~\cite{bandt2023two}. However, they remain unexplored for graph signals. Our contributions are:
\begin{itemize}[leftmargin=*, nosep]
    \item The first definition of permutation patterns for graph signals as a way to characterise them at granular level.
    \item Extension of the contrasts $\alpha$ (turning rate) and $\beta$ (up-down balance) to graph signals for detailed pattern analysis. 
    \item The study of the behaviour of $\alpha$ and $\beta$ for synthetic benchmarks of graph signals.
    \item The illustration of graph permutation patterns to characterise local changes in neuroimaging datasets in mild cognitive impairment, a prodromal phase of Alzheimer's disease.
\end{itemize}

\section{Graph-based Permutation patterns}  

\subsection{Notation}
Let \( G = (\V,\E,\A) \) represent a \emph{simple undirected graph} with vertex set \( \V = \{v_1, v_2, \dots, v_N\} \) and edge set \( \E \) defined as \( \E \subset \{(v_i,v_j) | v_i,v_j \in \V\} \). The adjacency matrix \( \A \) is an \( N \times N \) symmetric matrix with \( \A_{i j} = 1 \) if an edge connects \( v_i \) and \( v_j \), and \( \A_{i j} = 0 \) otherwise.

A \emph{graph signal} \( \X \) maps \( \V \to \R \).  \( \X = [x_1, x_2, \dots, x_N]^T \) is a column vector where the indices correspond to \( \V \).

A \emph{permutation} \(\pi\) is a bijection \(\pi : \Nm \to \Nm\) with \(\Nm = \{1, 2, \dots, m\}\). Using shorthand, \(\pi_k\) stands for \(\pi(k)\) for each \(k \in \Nm\), and the permutation is expressed as \(\pi = \pi_1\pi_2\dots \pi_m\). The complete set of permutation patterns is denoted by \(\Pi\). For instance, the notation \(\pi = 321\) implies \(\pi(1) = 3\), \(\pi(2) = 2\), and \(\pi(3) = 1\). Given a vector \(\textbf{x} = (x_1, x_2, \dots, x_m) \in \R^m\), \(\textbf{x}\) is said to exhibit the pattern \(\pi\) if \( \pi_i < \pi_j \) is true if and only if \( x_i < x_j \).
Lastly, \(\card{\cdot}\) denotes cardinality.

\subsection{Graph-based permutation patterns}\label{sec:alg}
Let $\X$ be a graph signal defined on $\G$, and $2\leq m\in\Na$ be the \emph{embedding dimension}. The graph-based permutation patterns are defined as follows: 
\setlength{\abovedisplayskip}{3pt}
\setlength{\belowdisplayskip}{3pt}
\begin{enumerate}[leftmargin=*, nosep]
	\item The \emph{embedding matrix} $\textbf{Y}\in\R^{N\times m}$ is given by
	$\textbf{Y}=[\textbf{y}_0,  \textbf{y}_1,  \cdots,  \textbf{y}_{m-1}]$, defined by
	\begin{equation}\label{eq:emb_mat}
		\textbf{y}_k=D^{k}\A^{k} \textbf{X}\in \R^{N\times 1}\;, \quad k=0,1,\dots,m-1\;,
	\end{equation}
	where $D^{k}$ is the diagonal matrix $D^{k}_{ii}=1/\sum_{j=1}^N (\A^{k})_{ij}$. 
	\item \emph{Graph-based permutation patterns.} Each vertex of the graph is assigned an \emph{embedding vector} and mapped to a unique permutation pattern.
	Formally, the \emph{embedding vectors} consist of $m$ numbers corresponding to each row of the matrix $\Y$, i.e., $\row_i(\Y)=\left( y_{ij}\right)_{j=1}^m $ for $i=1,2,\dots,N$. 
	Each embedding vector (one for each vertex of the graph) is uniquely mapped to a permutation pattern, i.e.,  $v_i\rightarrow\row_i(\Y)\rightarrow \pi\in \Pi$.
 	\item \emph{Relative frequencies.} For each dispersion pattern $\pi\in\Pi$, its relative frequency, $\rho\left(\pi\right)\in [0,1]$, is obtained as:
	\begin{equation}\label{eq:probability}
  \rho\left(\pi\right)=\card{\left\{v_i \mid v_i \in \V \text { and } v_i \text { has type } \pi\right\}} / N .
	\end{equation}
\end{enumerate}
 
\subsection{Permutation patterns and contrast for length \( 3 \)}
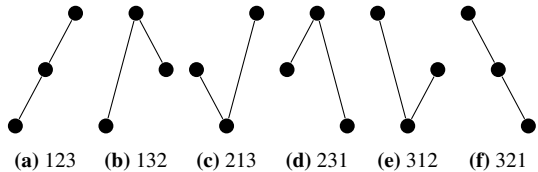
\begin{figure}[htb!]
	\centering
	\begin{subfigure}{0.14\linewidth}
		\centering
		\begin{tikzpicture}[xscale=0.4, yscale=.75]
			\node[fill=black, circle, inner sep=2pt] (a) at (1,1) {};
			\node[fill=black, circle, inner sep=2pt] (b) at (2,2) {};
			\node[fill=black, circle, inner sep=2pt] (c) at (3,3) {};
			\draw (a) -- (b) -- (c);
		\end{tikzpicture}
		\caption{123}
	\end{subfigure}%
	\begin{subfigure}{0.14\linewidth}
		\centering
		\begin{tikzpicture}[xscale=0.4, yscale=.75]
			\node[fill=black, circle, inner sep=2pt] (a) at (1,1) {};
			\node[fill=black, circle, inner sep=2pt] (b) at (2,3) {};
			\node[fill=black, circle, inner sep=2pt] (c) at (3,2) {};
			\draw (a) -- (b) -- (c);
		\end{tikzpicture}
		\caption{132}
	\end{subfigure}%
	\begin{subfigure}{0.14\linewidth}
		\centering
		\begin{tikzpicture}[xscale=0.4, yscale=.75]
			\node[fill=black, circle, inner sep=2pt] (a) at (1,2) {};
			\node[fill=black, circle, inner sep=2pt] (b) at (2,1) {};
			\node[fill=black, circle, inner sep=2pt] (c) at (3,3) {};
			\draw (a) -- (b) -- (c);
		\end{tikzpicture}
		\caption{213}
	\end{subfigure}%
	\begin{subfigure}{0.14\linewidth}
		\centering
		\begin{tikzpicture}[xscale=0.4, yscale=.75]
			\node[fill=black, circle, inner sep=2pt] (a) at (1,2) {};
			\node[fill=black, circle, inner sep=2pt] (b) at (2,3) {};
			\node[fill=black, circle, inner sep=2pt] (c) at (3,1) {};
			\draw (a) -- (b) -- (c);
		\end{tikzpicture}
		\caption{231}
	\end{subfigure}%
	\begin{subfigure}{0.14\linewidth}
		\centering
		\begin{tikzpicture}[xscale=0.4, yscale=.75]
			\node[fill=black, circle, inner sep=2pt] (a) at (1,3) {};
			\node[fill=black, circle, inner sep=2pt] (b) at (2,1) {};
			\node[fill=black, circle, inner sep=2pt] (c) at (3,2) {};
			\draw (a) -- (b) -- (c);
		\end{tikzpicture}
		\caption{312}
	\end{subfigure}%
	\begin{subfigure}{0.14\linewidth}
		\centering
		\begin{tikzpicture}[xscale=0.4, yscale=.75]
			\node[fill=black, circle, inner sep=2pt] (a) at (1,3) {};
			\node[fill=black, circle, inner sep=2pt] (b) at (2,2) {};
			\node[fill=black, circle, inner sep=2pt] (c) at (3,1) {};
			\draw (a) -- (b) -- (c);
		\end{tikzpicture}
		\caption{321}
	\end{subfigure}%
	\caption{The six permutation patterns for \( m = 3 \)}
    \label{fig:patternsm3}
\end{figure}

Here, we focus on \( m = 3 \) (depicted in Fig.~\ref{fig:patternsm3}), which is a well-studied case in univariate time series \cite{Bandt2023}. Increased pattern lengths make statistical estimates of pattern frequencies less accurate and their interpretation increasingly challenging~\cite{Bandt2023}.

The turning rate, denoted as \(\alpha\), quantifies the prevalence of turning points relative to monotonically increasing or decreasing segments within a time series. The up-down balance, denoted as \(\beta\), distinguishes upward and downward patterns~\cite{Bandt2023}. They are traditionally defined as:
\begin{eqnarray}
\alpha &=& \rho(132) + \rho(213) + \rho(231) + \rho(312)\;;\\
\beta  &=& \rho(123) - \rho(321)\;.
\end{eqnarray}
As detailed in Sec.~\ref{sec:alg}, we can expand the traditional definitions of \(\alpha\) and \(\beta\) beyond their original scope in~\cite{Bandt2023}. Notably, the graph signal contrast matches that of a univariate time series when the graph is a directed path. However, our graph-based approach allows us to craft more encompassing contrasts, integrating both graph's topology and data.

A critical nuance of our methodology is its ability to assign a distinct pattern to each sample. This contrasts with time series permutation patterns. This granularity affords deeper insights into graph signals, enabling precise characterization of each data point.
 
\section{Benchmarking on synthetic data}
\textbf{$\MIX$ Processing.} In a \emph{Random Geometric Graph} ($\RGG$), each vertex \( v_i \in \mathcal{V} \) is assigned a random 2D coordinate \( \mathbf{z}_i=( z_i^1, z_i^2) \in [0,1]^2 \). Vertices \( v_i \) and \( v_j \), are connected if their coordinates' distance is  \(\leq r \). For vertex \( v_i \), the signal value is determined by:
\begin{equation}
  \MIX(v_i) = \left( (1-R)S(\mathbf{z}_i) + RW(\mathbf{z}_i) \right),  1 \leq i \leq N\;.  
\end{equation}
Here, $R$ is a random variable with a probability \( p \) of being 1 and \( 1-p \) of being 0, $W$ represents uniformly distributed white noise, and \( S(\mathbf{z_i})= \sin(2\pi f z_i^1) + \sin(2\pi f z_i^2) \). For \( p=0 \), $\MIX(v_i)=S(\mathbf{z}_i)$, which is a regular periodic signal and for \( p=1 \), $\MIX$ is entirely noise, allowing exploration of both structured and stochastic graph behaviors as in~\cite{carrasco2023dispersion}.

\begin{figure}[htb!]
\centering
\includegraphics[width=7cm]{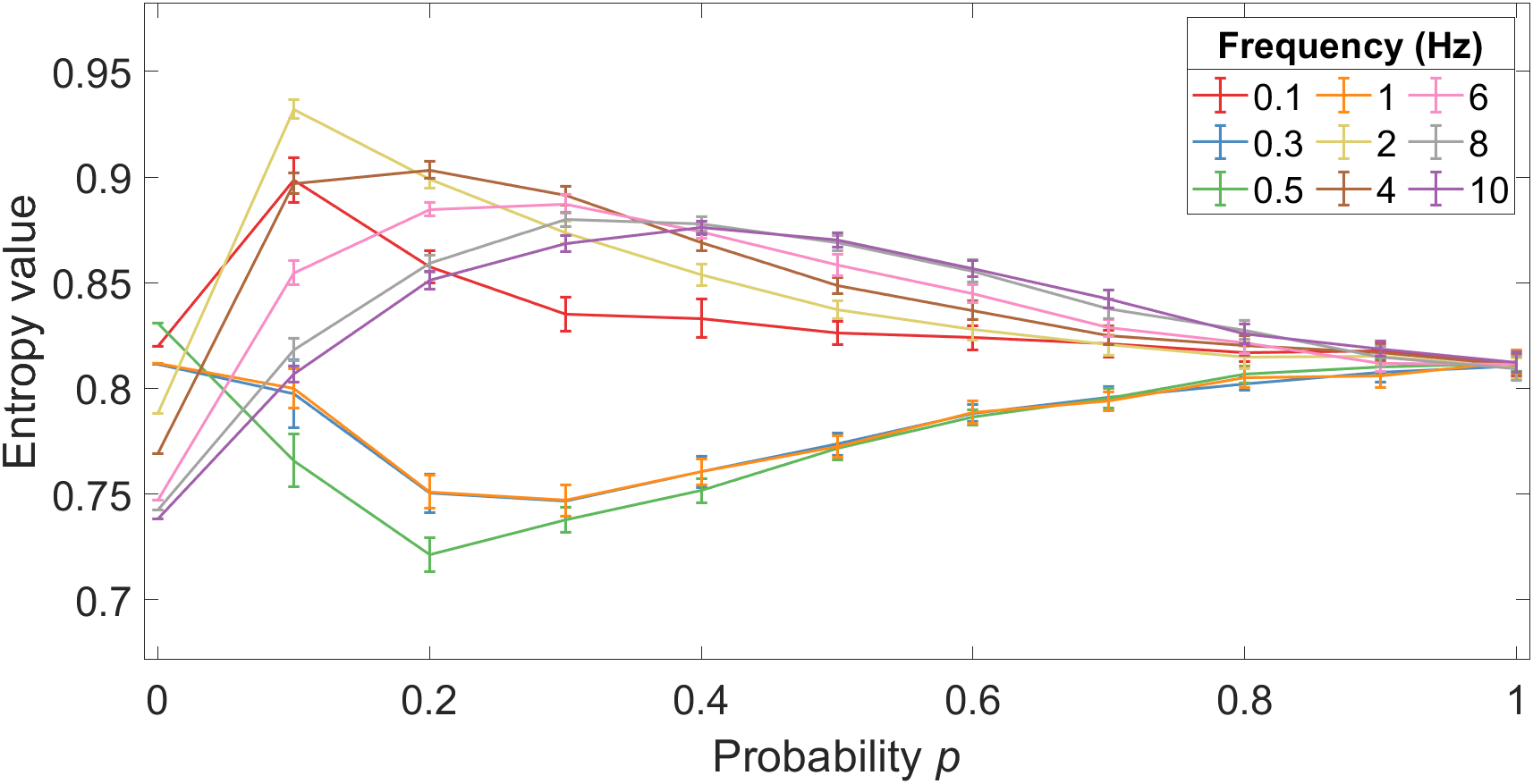}
\caption{\small Mean and std of values of $\PEG$ for a consistent graph across increasing noise levels and varying frequencies.}
\label{fig:freq_pPEG}
\end{figure}

\textbf{Permutation Entropy Analysis.} Our investigation centered on discerning the irregularities of the $\MIX$ graph signal, especially those affected by the variations in parameters $f$ and $p$. Throughout this process, the $\RGG$ parameters remained constant at $N=1500$ and $r=0.06$, and the study spanned 20 realizations. The initial step involved computing the entropy -- $\PEG$ -- for the graph signal values. The resulting entropy mean and standard deviation (std), evaluated across different $\MIX$ process frequencies, are presented in Fig.~\ref{fig:freq_pPEG}. Our observations indicate that relying purely on the permutation entropy value falls short in delivering clear insights. Specifically, this method does not effectively track the signal dynamics amid rising noise or shifting frequency.

\textbf{Permutation Pattern Analysis}
Using our graph-based permutation pattern analysis, we present an exploration of the $\MIX$ process. 

\textbf{Baseline Behavior at \( p=0 \):} At this level, the $\MIX$ signal naturally shows periodic tendencies. When the frequency increases, \( \alpha \) slightly drops due to fewer local extrema. As expected, \( \beta \approx 0 \) because the $\MIX$ signal for $p=0$ is governed by the sinusoidal components, resulting in a similar number of monotonically increasing and decreasing patterns. 

\begin{figure}[htb!]
 	\begin{minipage}[b]{.49\linewidth}
 		\centering
 		\centerline{\includegraphics[width=4.2cm]{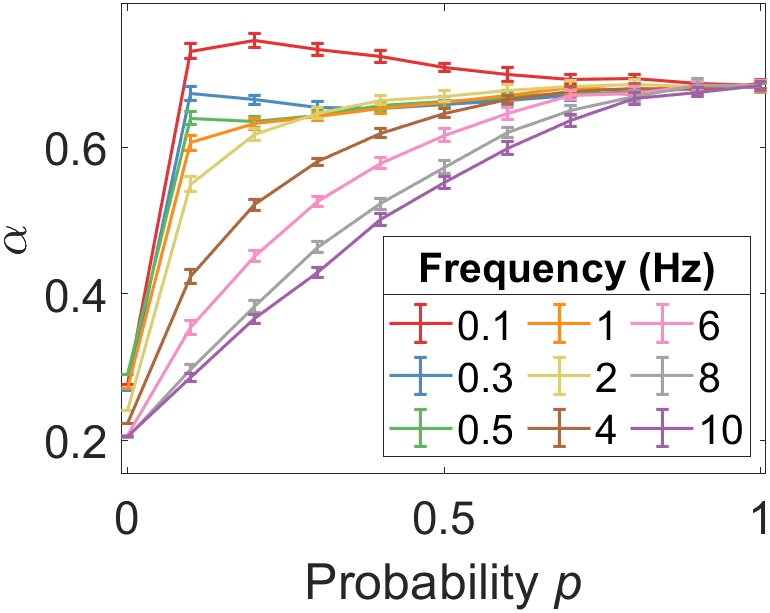}}
 		\centerline{\small (a) Turning rate $\alpha$}\medskip
 	\end{minipage}
 	\hfill
 	\begin{minipage}[b]{0.49\linewidth}
 		\centering
 		\centerline{\includegraphics[width=4.2cm]{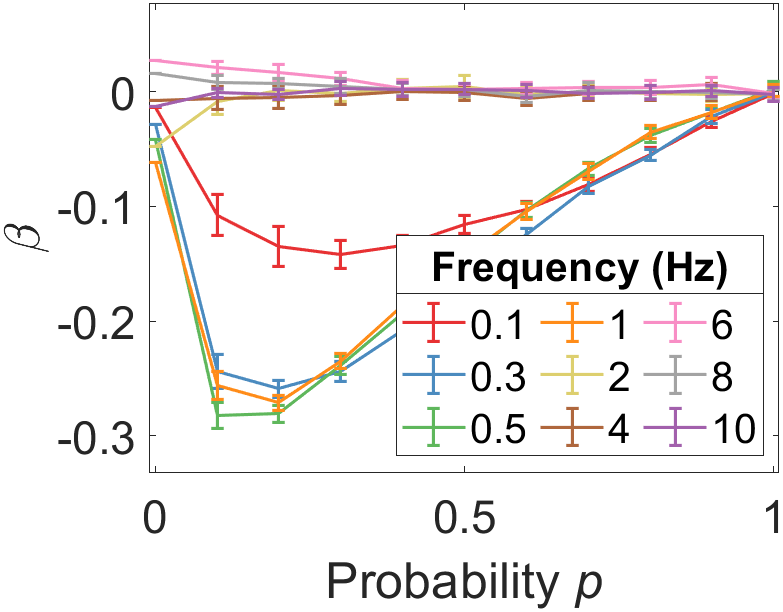}}
 		\centerline{\small (b) Up-down balance $\beta$}\medskip
 	\end{minipage}
 	\caption{Graph-based contrasts for the $\MIX$ process.}
 	\label{fig:freq_pPEG_2}
 \end{figure}

\textbf{Effects of Noise and \( p \):} An increase in noise or the \( p \) parameter leads to a rise in \( \alpha \), reflecting a decrease in the number of strictly monotonous patterns. Notably, a slight change in \( p \) from \( 0 \) to \( 0.1 \) causes a significant increase in \( \alpha \). This underscores the $\MIX$ process's sensitivity to small changes, with the trend stabilizing for higher \( p \) values, as illustrated in Fig.~\ref{fig:freq_pPEG_2}(a). The interplay between the noise and period of the sinusoids results in values of $\beta$ deviating from 0.

\textbf{Frequency Relationship:} Notably, for a constant \( p \), \( \alpha \) shows a direct relationship with frequency: a decrease in frequency leads to a heightened \( \alpha \). Conversely, higher frequencies result in fewer local points, leading to a reduced \( \alpha \).

Increases in \( r \) augments graph connectivity, thereby enhancing sensitivity to frequency changes.

\textbf{The \( \beta \)-\( \alpha \) Complementarity:} \( \alpha \) and \( \beta \) display different behaviours, confirming that they both provide complementary information, as shown in  Fig.~\ref{fig:freq_pPEG_2}.

\section{Real-world illustration in MCI}
Dementia currently affects over 50 million people worldwide and is expected to triple by 2050~\cite{Scheltens2021}. Alzheimer's disease (AD) is the main cause of dementia and causes immense emotional and financial strain on families and healthcare services. Its early stages are often categorized by stages of Mild Cognitive Impairment (MCI), often progressing (within 4 years) to the dementia stage of AD~\cite{Scheltens2021}. To understand this progression, we explore a novel MRI model of AD and its potential use in characterizing the stages of disease.

\subsection{Participants and task}
Participants from the longitudinal study ~\cite{Parra2022} were assessed with a battery of neuropsychological tests commonly used to assess dementia, grouping subjects into early Mild Cognitive Impairment (eMCI), MCI, and Alzheimer's disease converters after a 2-year follow up (MCIc)~\cite{Parra2022}. From these, 8 healthy controls (Age: $76.50\pm5.21$, Sex: 2M; 6F), 7 eMCI (Age: $76.86\pm6.41$, Sex: 4M; 3F), 10 MCI (Age: $72.30\pm5.64$, Sex: 5M; 5F), and 6 MCIc subjects (Age: $76.33\pm5.09$, Sex: 4M; 2F) were selected to undergo DTI and fMRI acquisition during which they performed a Visual Short-Term Memory Binding Task (VSTMBT).

The VSTMBT~\cite{Parra2010} is a task sensitive to memory related changes in early stage AD. Participants were presented non-nameable coloured shapes on a screen for 2s (encoding). They must memorize this information after a blank screen is shown for a variable amount of time of 2, 4, 6, or 8s (maintenance). Then, they are presented the same or a different set of associations of shapes and colours for 4s. The participants must determine if they are the same or different (probe), followed by an inter-trial interval before repetition. In this study, we focus on the encoding phase of the task to assess the formation of memories in healthy and diseased groups. 

\subsection{Graph and signal construction}

fMRI data was collected with a GE Signa Horizon HDxt 1.5T clinical scanner. During the VSTMBT, contiguous interleaved axial gradient EPI were collected alongside the intercommissural plane throughout two continuous runs ($\text{TR/TE} = 2000/40\text{ms}$; $\text{matrix} = 64 \times 64$; $\text{fov} = 24\text{cm}$; $\text{thickness} = 5\text{mm}$; $\text{gap} = 0\text{mm}$). 

Outlier detection, realignment, slice-timing correction, co-registration of the structural ($T_1$) and functional images to the MNI space, segmentation, and normalization were performed with SPM12. ROIs for each subject are defined using an 85 region atlas, detailed below. For each ROI, the mean signal is acquired across the voxels in that region and highpass filtered (0.06Hz) to avoid fMRI signal drift. 

For Diffusion MRI, 3 $T_2$-weighted ($\text{b} = 0 \text{s mm}^{-2}$) and sets of diffusion-weighted ($b = 1000\text{s mm}^{-2}$) single-shot spin-echo-planar (EP) volumes were acquired with diffusion gradients applied in 32 non-collinear directions. Subsequent volumes were in the axial plane ($\text{fov} = 240 \times 240$; $\text{matrix} = 128 \times 128$; $ \text{thickness} = 2.5\text{mm}$), giving voxel dimensions of $1.875 \times 1.875 \times 2.5\text{mm}$. 

A $T_1$ weighted volume was also acquired with 1.3 mm$^3$ voxel dimensions. This volume was parcellated into 85 ROIs with the Desikan-Killiany atlas combined with additional regions acquired via sub-cortical segmentation detailed in~\cite{buchanan2014test}, and the brain-stem using FreeSurfer. Standard pre-processing was applied following~\cite{buchanan2014test}, resulting in DTI networks where edge weight was determined by the streamline density (SD) between regions, corrected for ROI size.  

\subsection{Results}
\begin{figure}[htb!]
\centering
	\begin{minipage}[b]{0.25\textwidth}
		\includegraphics[width=\linewidth]{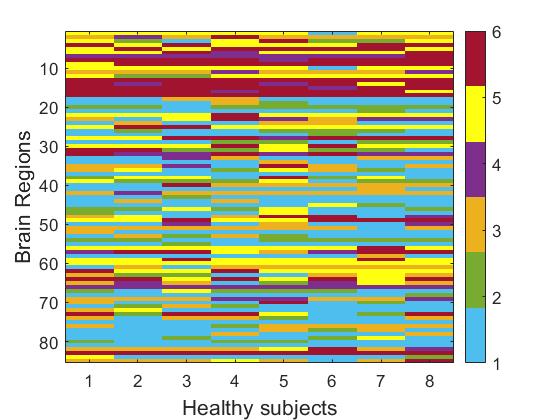}
		\centerline{\small (a) Patterns across subjects}
	\end{minipage}
	\hspace{.4cm}
	\begin{minipage}[b]{0.15\textwidth}
		\includegraphics[width=\linewidth]{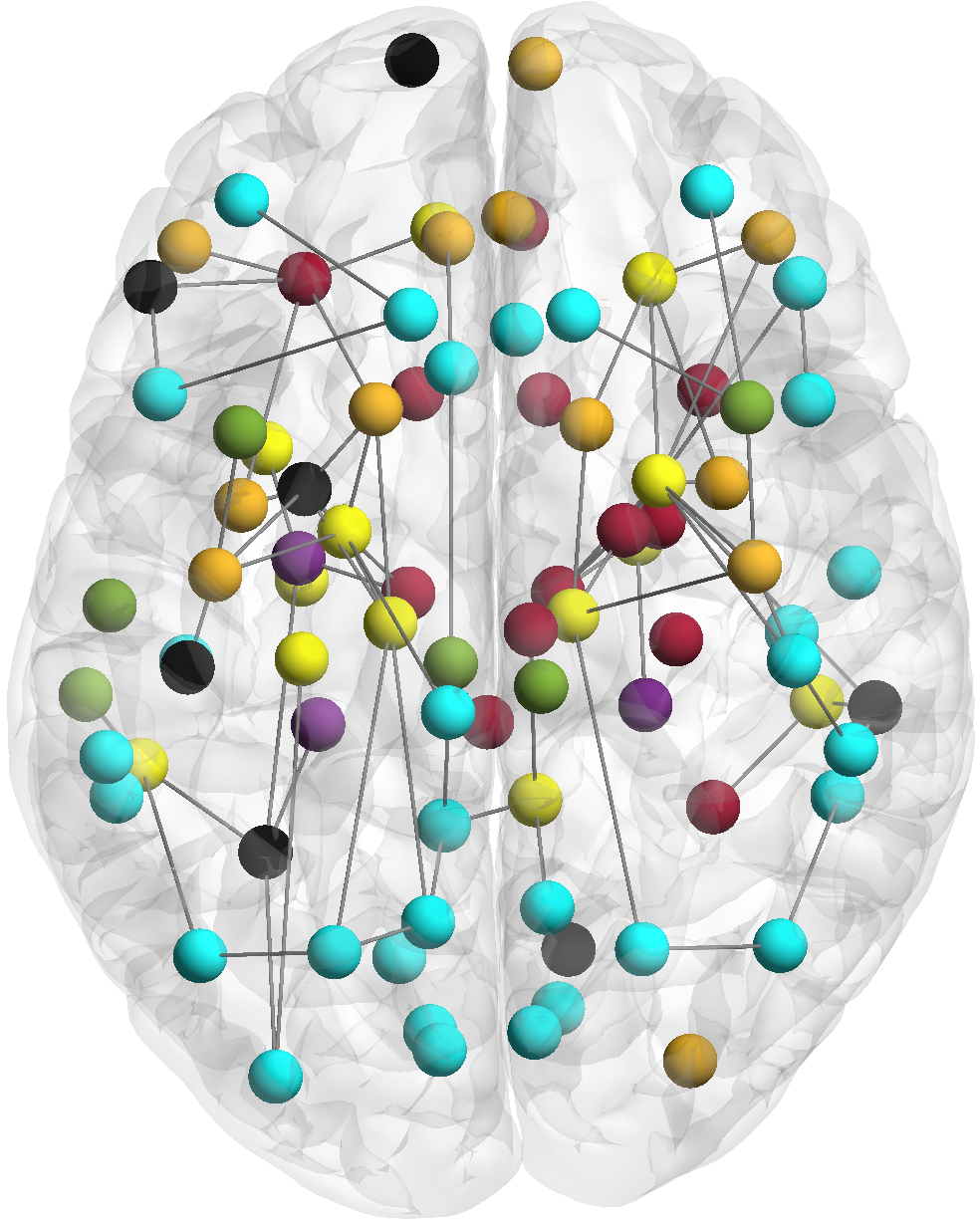}
		\centerline{\small (b) Brain visualization}
	\end{minipage}
	\caption{(a) visualizes the distribution of patterns (rows), across subjects (columns). In (b), patterns for a node were based on the mode of the distribution of patterns for the healthy group, but only when that pattern was in at least half of subjects (black otherwise). (b) was generated with the BrainNet viewer tool~\cite{xia2013brainnet}.}
	\label{fig:healthypatterns}
\end{figure}

We calculate the graph-based patterns as per Sec.~\ref{sec:alg} with $m=3$. The graph is the subject's SD-weighted DTI network and the signal \textit{at each node} is the mean signal across the encoding phases of the task, yielding a pattern at each node. Though limited by sample size, in the healthy brain networks, we observe the existence of dominant patterns in some clusters, such as patterns 5\&6 in ROIs 1-18, and patterns 1\&2 in ROIs 75-81 (see Fig.~\ref{fig:healthypatterns}a), suggesting that there may be some identifying patterns associated with the encoding phase of the VSTMBT. (Here we refer to patterns \#1 to \#6 following the order as in Fig.~\ref{fig:patternsm3}.)

To determine whether patterns change with disease, we perform chi-squared analysis comparing the per node patterns between each pair of control and disease groups. Furthermore, we assess the stability of the resulting $p$-value by permuting the control and disease groups 1000 times to calculate how often our original $p$-value ($p$) is smaller than that of the randomly permuted groups ($p'$). Due to the limited sample size, we took a conservative approach to report regions where both $p, p'\leq 0.05$. 

\begin{table}[htb!]
\small
\centering 
\begin{tabularx}{.91\linewidth}{@{}p{1.5cm}p{3.25cm}p{1cm}p{.9cm}}
\textbf{Control vs.} & \textbf{ROIs}              & \textbf{$p$-value} & \textbf{$p<p'$} \\ \hline
\textbf{eMCI}        & Right-lateralorbitofrontal & 0.019            & 0.009                   \\ \hline
\textbf{MCI}         & Right-entorhinal           & 0.015            & 0.002                   \\
                     & Right-lateralorbitofrontal & 0.020            & 0.027                   \\
                     & Right-parahippocampal      & 0.010            & 0                       \\ \hline
\textbf{MCIc}        & Left-hippocampus           & 0.049            & 0.050                   \\
                     & Left-caudalmiddlefrontal   & 0.036            & 0.033                   \\
                     & Left-medialorbitofrontal   & 0.031            & 0.008                   \\
                     & Right-lateralorbitofrontal & 0.005            & 0                       \\
                     & Right-paracentral          & 0.049            & 0.021                                  
\end{tabularx}
\caption{Statistical tests to find regions with significant  differences in the distribution of graph-based permutation patterns between control and different stages of MCI.}
\label{tab:chi-table}
\end{table}

We find that, as the disease progresses (Table~\ref{tab:chi-table}), the number of regions which exhibit a significant change in pattern increases, following a neuroanatomical trajectory consistent with that described by the AD continuum, i.e., Medial Temporal Lobe (MTL) regions first and then broader impact including frontal lobes~\cite{bastin2023targeting, didic2011memory, Parra2022}. Not only was the gross neuroanatomical spread of AD pathology found, but our method identified the more fine grained distribution of pathology within the MTL characterizing the earliest stages of AD (i.e., entorhinal which feeds to parahippocampal and hippocampal regions~\cite{bastin2023targeting, didic2011memory, Parra2022}). Specifically, we see an increasing change in the orbitofrontal cortex as the disease progresses both with decreasing $p$-value in the Right lateral orbitofrontal, along with the presence of the medial orbitofrontal at the later stage of disease. The orbitofrontal cortex is a vulnerable region to early deposition of amyloid plaques, a key bio-marker in AD progression~\cite{sepulcre2013vivo,Resnick2007}. Similarly, damage in the entorhinal, paracentral, frontal, and hippocampal structures are other early indicators of AD in studies of amyloid deposition and structural and functional MRI~\cite{Fjell2014, Laakso2000, Zhao2018}.

\begin{figure}[htb!]
\centering
	\begin{minipage}[b]{0.15\textwidth}
		\centering
		\includegraphics[width=\linewidth]{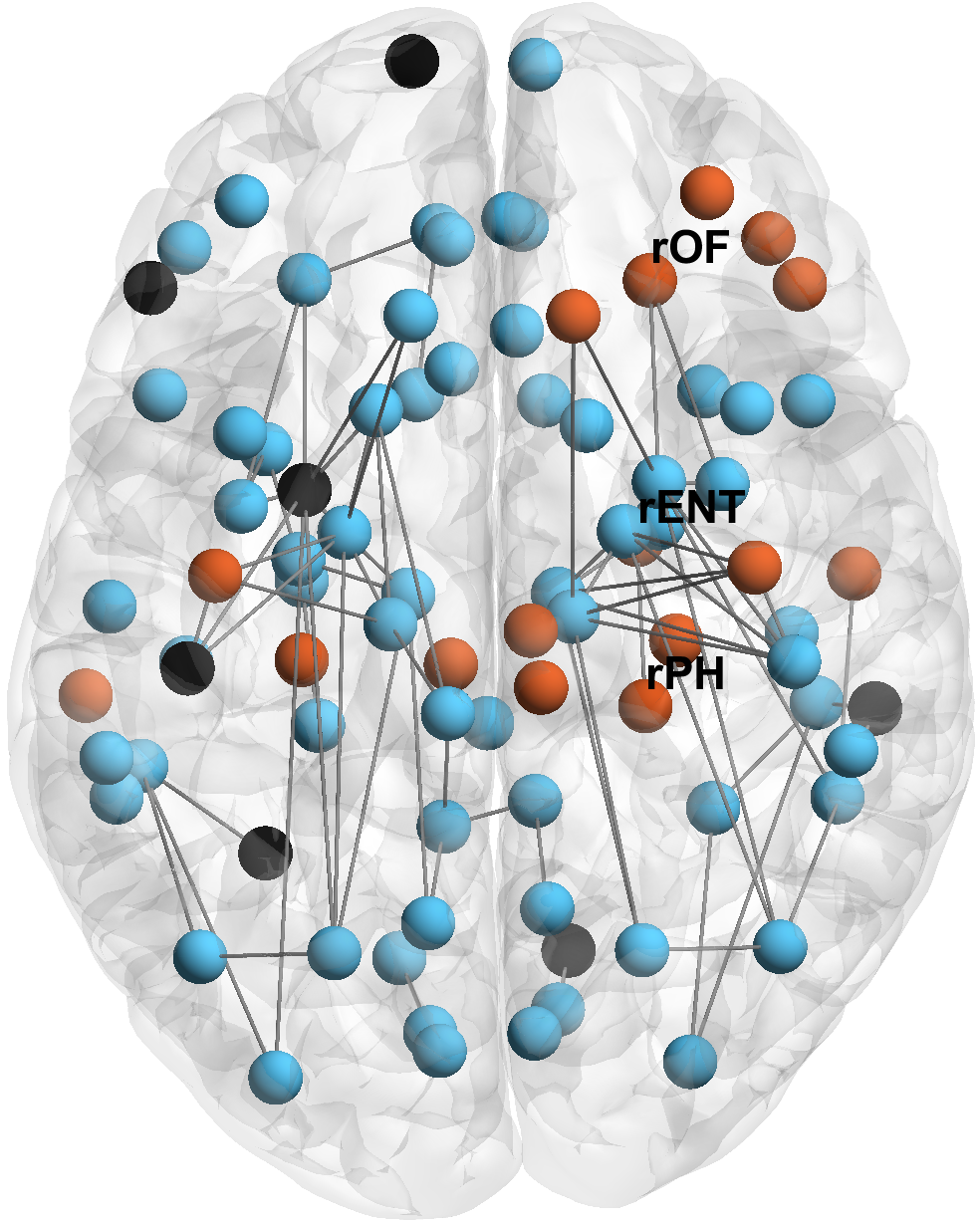}
		\centerline{\small (a) Control vs. MCI}\medskip
	\end{minipage}
	\hspace{.8cm}
	\begin{minipage}[b]{0.15\textwidth}
		\includegraphics[width=\linewidth]{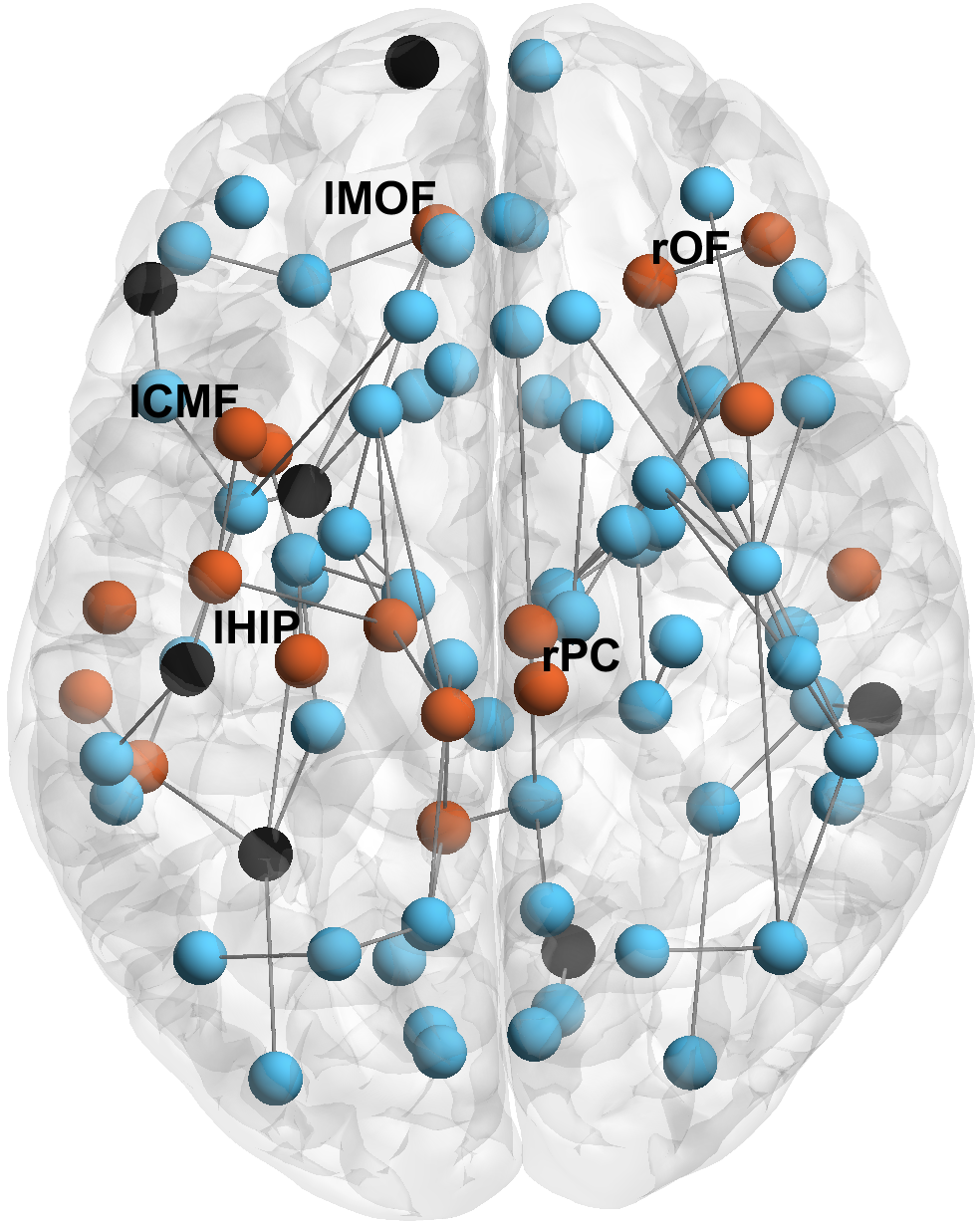}
		\centerline{\small (b) Control vs. MCIc}\medskip
	\end{minipage}
	\caption{Changes in pattern between healthy and disease. Note that only $2\%$ of DTI edges are drawn for clarity. Generated with the BrainNet viewer tool~\cite{xia2013brainnet}.}
	\label{fig:patternschange}
\end{figure}

Additionally, we look at pattern frequency changes between groups at granular scale. Namely, we identify the most dominant pattern per node for each subject group that appears in at least half of the subjects. This is visualized in Fig.~\ref{fig:patternschange}. Here, nodes in orange are those that have changed pattern, blue indicated no change, black had no definitive pattern within the control group, and labelled nodes are from Table~\ref{tab:chi-table}.

\section{Conclusion and limitations}
We have extended the analysis of permutation patterns to graph signals, providing a novel lens to view and analyze such data at granular scale. Our findings indicate that the turning rate ($\alpha$) and up-down balance ($\beta$) serve as effective tools for graph-based pattern analysis. Furthermore, we identify the potential use of graph based permutation patterns for multi-modal MRI data of MCI. Though limited by sample size, our results motivate larger studies of graph based permutation patterns on other real-world data such as MRI-based brain graph signals. 

\bibliographystyle{abbrv}
\bibliography{main}

\begin{thebibliography}{1}

\bibitem{Bandt2002}
Christoph Bandt and Bernd Pompe,
\newblock ``{Permutation Entropy: A Natural Complexity Measure for Time
  Series},''
\newblock {\em Physical Review Letters}, vol. 88, no. 17, pp. 174102, apr 2002.

\bibitem{Morel2021}
Cristina Morel and Anne Humeau-Heurtier,
\newblock ``{Multiscale permutation entropy for two-dimensional patterns},''
\newblock {\em Pattern Recognition Letters}, vol. 150, pp. 139--146, oct 2021.

\bibitem{Zunino2009}
Luciano Zunino, Massimiliano Zanin, Benjamin~M Tabak, Dar{\'{i}}o~G
  P{\'{e}}rez, and Osvaldo~A Rosso,
\newblock ``{Forbidden patterns, permutation entropy and stock market
  inefficiency},''
\newblock {\em Physica A}, vol. 388, pp. 2854--2864, 2009.

\bibitem{Bandt2023}
Christoph Bandt,
\newblock ``{Statistics and contrasts of order patterns in univariate time
  series},''
\newblock {\em Chaos}, vol. 33, no. 3, 2023.

\bibitem{fabila2022permutation}
John~Stewart Fabila-Carrasco, Chao Tan, and Javier Escudero,
\newblock ``Permutation entropy for graph signals,''
\newblock {\em IEEE Transactions on Signal and Information Processing over
  Networks}, vol. 8, pp. 288--300, 2022.

\end{thebibliography}


\begin{thebibliography}{10}

\bibitem{azami2016improved}
H.~Azami and J.~Escudero.
\newblock Improved multiscale permutation entropy for biomedical signal
  analysis: Interpretation and application to electroencephalogram recordings.
\newblock {\em Biomedical Signal Processing and Control}, 23:28--41, 2016.

\bibitem{azami2023entropy}
H.~Azami, L.~Faes, J.~Escudero, et~al.
\newblock Entropy analysis of univariate biomedical signals: Review and
  comparison of methods.
\newblock {\em Frontiers in Entropy across the Disciplines: Panorama of
  Entropy: Theory, Computation, and Applications}, pages 233--286, 2023.

\bibitem{Bandt2023}
C.~Bandt.
\newblock {Statistics and contrasts of order patterns in univariate time
  series}.
\newblock {\em Chaos}, 33(3), 2023.

\bibitem{Bandt2002}
C.~Bandt and B.~Pompe.
\newblock {Permutation Entropy: A Natural Complexity Measure for Time Series}.
\newblock {\em Physical Review Letters}, 88(17):174102, apr 2002.

\bibitem{bandt2023two}
C.~Bandt and K.~Wittfeld.
\newblock Two new parameters for the ordinal analysis of images.
\newblock {\em Chaos: An Interdisciplinary Journal of Nonlinear Science},
  33(4), 2023.

\bibitem{bastin2023targeting}
C.~Bastin and E.~Delhaye.
\newblock Targeting the function of the transentorhinal cortex to identify
  early cognitive markers of alzheimer’s disease.
\newblock {\em Cognitive, Affective, \& Behavioral Neuroscience}, pages 1--11,
  2023.

\bibitem{bessadok2022graph}
A.~Bessadok, M.~A. Mahjoub, and I.~Rekik.
\newblock Graph neural networks in network neuroscience.
\newblock {\em IEEE Transactions on Pattern Analysis and Machine Intelligence},
  45(5):5833--5848, 2022.

\bibitem{buchanan2014test}
C.~R. Buchanan, C.~R. Pernet, et~al.
\newblock Test--retest reliability of structural brain networks from diffusion
  mri.
\newblock {\em Neuroimage}, 86:231--243, 2014.

\bibitem{cao2004detecting}
Y.~Cao, W.-w. Tung, et~al.
\newblock Detecting dynamical changes in time series using the permutation
  entropy.
\newblock {\em Physical review E}, 70(4):046217, 2004.

\bibitem{didic2011memory}
M.~Didic, E.~J. Barbeau, O.~Felician, E.~Tramoni, E.~Guedj, M.~Poncet, and
  M.~Ceccaldi.
\newblock Which memory system is impaired first in alzheimer's disease?
\newblock {\em Journal of Alzheimer's Disease}, 27(1):11--22, 2011.

\bibitem{dong2020graph}
X.~Dong, D.~Thanou, L.~Toni, M.~Bronstein, and P.~Frossard.
\newblock Graph signal processing for machine learning: A review and new
  perspectives.
\newblock {\em IEEE Signal processing magazine}, 37(6):117--127, 2020.

\bibitem{fabila2022permutation}
J.~S. Fabila-Carrasco, C.~Tan, and J.~Escudero.
\newblock Permutation entropy for graph signals.
\newblock {\em IEEE Transactions on Signal and Information Processing over
  Networks}, 8:288--300, 2022.

\bibitem{carrasco2023dispersion}
J.~S. Fabila-Carrasco, C.~Tan, and J.~Escudero.
\newblock Dispersion entropy for graph signals.
\newblock {\em Chaos, Solitons and Fractals}, 2023.

\bibitem{Fjell2014}
A.~M. Fjell, L.~McEvoy, et~al.
\newblock {What is normal in normal aging? Effects of aging, amyloid and
  Alzheimer's disease on the cerebral cortex and the hippocampus}.
\newblock {\em Progress in Neurobiology}, 117:20--40, jun 2014.

\bibitem{huang2018graph}
W.~Huang, T.~A. Bolton, et~al.
\newblock A graph signal processing perspective on functional brain imaging.
\newblock {\em Proceedings of the IEEE}, 106(5):868--885, 2018.

\bibitem{Laakso2000}
M.~P. Laakso, G.~B. Frisoni, et~al.
\newblock {Hippocampus and entorhinal cortex in frontotemporal dementia and
  Alzheimer's disease: a morphometric MRI study}.
\newblock {\em Biological Psychiatry}, 47(12):1056--1063, jun 2000.

\bibitem{li2021graph}
R.~Li, X.~Yuan, M.~Radfar, P.~Marendy, W.~Ni, T.~J. O’Brien, and P.~M.
  Casillas-Espinosa.
\newblock Graph signal processing, graph neural network and graph learning on
  biological data: a systematic review.
\newblock {\em IEEE Reviews in Biomedical Engineering}, 16:109--135, 2021.

\bibitem{morabito2012multivariate}
F.~C. Morabito, D.~Labate, F.~L. Foresta, A.~Bramanti, et~al.
\newblock Multivariate multi-scale permutation entropy for complexity analysis
  of alzheimer’s disease eeg.
\newblock {\em Entropy}, 14(7):1186--1202, 2012.

\bibitem{Morel2021}
C.~Morel and A.~Humeau-Heurtier.
\newblock {Multiscale permutation entropy for two-dimensional patterns}.
\newblock {\em Pattern Recognition Letters}, 150:139--146, oct 2021.

\bibitem{ortega2022introduction}
A.~Ortega.
\newblock {\em Introduction to graph signal processing}.
\newblock Cambridge University Press, 2022.

\bibitem{ortega2018graph}
A.~Ortega, P.~Frossard, et~al.
\newblock Graph signal processing: Overview, challenges, and applications.
\newblock {\em Proceedings of the IEEE}, 106(5):808--828, 2018.

\bibitem{Parra2010}
M.~A. Parra, S.~Abrahams, R.~H. Logie, et~al.
\newblock {Visual short-term memory binding deficits in familial Alzheimer's
  disease}.
\newblock {\em Brain}, 133(9):2702--2713, sep 2010.

\bibitem{Parra2022}
M.~A. Parra, C.~Calia, V.~Pattan, and S.~{Della Sala}.
\newblock {Memory markers in the continuum of the Alzheimer's clinical
  syndrome}.
\newblock {\em Alzheimer's Research and Therapy}, 14(1):1--16, dec 2022.

\bibitem{Resnick2007}
S.~M. Resnick, M.~Lamar, and I.~Driscoll.
\newblock {Vulnerability of the Orbitofrontal Cortex to Age-Associated
  Structural and Functional Brain Changes}.
\newblock {\em Annals of the New York Academy of Sciences}, 1121(1):562--575,
  dec 2007.

\bibitem{ribeiro2021entropy}
M.~Ribeiro, T.~Henriques, L.~Castro, et~al.
\newblock The entropy universe.
\newblock {\em Entropy}, 23(2):222, 2021.

\bibitem{riedl2013practical}
M.~Riedl, A.~M{\"u}ller, and N.~Wessel.
\newblock Practical considerations of permutation entropy: A tutorial review.
\newblock {\em The European Physical Journal Special Topics}, 222(2):249--262,
  2013.

\bibitem{Scheltens2021}
P.~Scheltens, B.~{De Strooper}, M.~Kivipelto, et~al.
\newblock {Alzheimer's disease}.
\newblock {\em The Lancet}, 397(10284):1577--1590, apr 2021.

\bibitem{sepulcre2013vivo}
J.~Sepulcre, M.~R. Sabuncu, A.~Becker, R.~Sperling, and K.~A. Johnson.
\newblock In vivo characterization of the early states of the amyloid-beta
  network.
\newblock {\em Brain}, 136(7):2239--2252, 2013.

\bibitem{xia2013brainnet}
M.~Xia, J.~Wang, and Y.~He.
\newblock Brainnet viewer: a network visualization tool for human brain
  connectomics.
\newblock {\em PloS one}, 8(7):e68910, 2013.

\bibitem{yin2014weighted}
Y.~Yin and P.~Shang.
\newblock Weighted multiscale permutation entropy of financial time series.
\newblock {\em Nonlinear Dynamics}, 78:2921--2939, 2014.

\bibitem{Zhao2018}
Q.~Zhao, H.~Lu, et~al.
\newblock {Evaluating functional connectivity of executive control network and
  frontoparietal network in Alzheimer's disease}.
\newblock {\em Brain Research}, 1678:262--272, jan 2018.

\bibitem{Zunino2009}
L.~Zunino, M.~Zanin, B.~M. Tabak, et~al.
\newblock {Forbidden patterns, permutation entropy and stock market
  inefficiency}.
\newblock {\em Physica A}, 388:2854--2864, 2009.

\end{thebibliography}
\end{document}